\documentclass[aps,prd,twocolumn,titlepage,nofootinbib,longbibrliography]{revtex4-1}
\usepackage{enumitem}
\usepackage{comment}
\usepackage{tabularx}
\usepackage{xcolor}
\usepackage{booktabs}
\usepackage{tabularx}
\usepackage{array}
\usepackage{tcolorbox}
\usepackage[utf8]{inputenc}
\tcbuselibrary{skins, breakable}
\usepackage{graphicx,physics}
\usepackage{color}
\usepackage{dcolumn}
\usepackage{latexsym}

\usepackage[normalem]{ulem}
\usepackage{hyperref,amssymb}
\usepackage{url}
\usepackage{color,soul}
\usepackage{graphicx}
\usepackage{verbatim}
\usepackage{multirow}
\usepackage[table]{xcolor}

\usepackage{amsmath}
\newcommand{\beq}{\begin{eqnarray}}
\newcommand{\eeq}{\end{eqnarray}}
\usepackage{mathrsfs}
\usepackage{float,soul}
\usepackage[dvipsnames]{xcolor}
\usepackage{mathtools}
\usepackage{slashed}
\usepackage{physics}	
\usepackage{graphicx}   
\usepackage{epstopdf}  
\usepackage{hyperref}   
\usepackage{bbold}
\usepackage{wasysym}
\usepackage{feynmp}
\usepackage{hyperref}
\usepackage{xcolor}
\hypersetup{colorlinks,
}
\usepackage{bm}
\usepackage{lineno}

\usepackage{booktabs}
\usepackage[table]{xcolor}

\usepackage{mdframed}
\usepackage{tikz}

\newcommand*\reddisk{\tikz{\fill[red,radius=0.8ex] (0,0) circle;}}
\newcommand*\bluedisk{\tikz{\fill[blue,radius=0.8ex] (0,0) circle;}}

\newmdenv[
  linecolor=gray!60,
  backgroundcolor=gray!5,
  linewidth=0.6pt,
  roundcorner=4pt,
  skipabove=10pt,
  skipbelow=10pt,
  innerleftmargin=5pt,
  innerrightmargin=5pt,
  innertopmargin=5pt,
  innerbottommargin=5pt
]{figframe}
\newtcolorbox{figbox}{
  colback=gray!5,
  colframe=gray!40!black,
  boxrule=0.6pt,
  arc=3pt,
  outer arc=3pt,
  boxsep=5pt,
  left=3pt,
  right=3pt,
  top=3pt,
  bottom=3pt,
  width=\textwidth,
  breakable,
}

\begin{document}

\title{\Large Topological Defects in Amorphous Solids}
\author{Matteo Baggioli$^{1}$}
\email{b.matteo@sjtu.edu.cn}
\author{Michael L. Falk$^{2,3,4,5}$}
\email{mfalk@jhu.edu}
\author{Walter Kob$^{6,7}$}
\email{walter.kob@umontpellier.fr}
\address{$^1$Wilczek Quantum Center, School of Physics and Astronomy, Shanghai Jiao Tong University, Shanghai 200240, China}
\address{$^2$Department of Material Sciences and Engineering, Johns Hopkins University, Baltimore, Maryland 21218, USA}
\address{$^3$Department of Mechanical Engineering, Johns Hopkins University, Baltimore, MD 21218, USA}
\address{$^4$Department of Physics and Astronomy, Johns Hopkins University, Baltimore, MD 21218, USA}
\address{$^5$Hopkins Extreme Materials Institute, Johns Hopkins University, Baltimore, MD 21218, USA}
\address{$^6$Department of Physics, University of Montpellier and CNRS, 34095 Montpellier, France}
\address{$^7$College of Physics, Chengdu University of Technology, Chengdu 610059, China}

\begin{abstract}
Topological defects (TDs) are crucial for understanding important physical properties of crystalline materials including mechanical failure, ion transport, and two-dimensional melting. This concept has not translated to disordered materials like glasses because these solids have no obvious reference structure that can be used to define TDs. As a result, key properties related to those listed above have typically been modeled using purely phenomenological approaches. Recent studies have demonstrated that certain observables commonly associated with TDs can also be identified in disordered solids indicating that topological concepts may be as crucial in amorphous solids as in crystals. This hints that TDs may offer a first-principles framework for understanding their mechanical response and complex spatiotemporal dynamics. In this Perspective, we review recent theoretical, numerical, and experimental studies that have exploited topological concepts to rationalize mechanical properties of amorphous solids. We also highlight pressing open questions and some promising directions for future research in the field. 
\end{abstract}
\maketitle

\section*{Goal of this perspective} 
In the last decade, condensed matter physics has seen a surge in the use of topological concepts in materials exhibiting long-range order since these afford insight into material features associated with defects~\cite{bhattacharjee2017topology,araujo2021topology,RevModPhys.82.3045,Wang2017,Bernevig2022}. In contrast, the application of such ideas to disordered materials have foundered. Very recent studies have reinvigorated the idea that topology can be usefully applied to elucidating the properties of amorphous materials such as glasses. The goal of this \textit{Perspective} is to review basic concepts underlying the definition of topological defects, to summarize recent applications to the structure and properties of glasses, and to raise open questions to guide the development of topological concepts in the context of disordered solids.

\section*{Topological defects and their fundamental role in physics}

The mathematics of topology concerns those properties of geometric objects that remain invariant under continuous transformations such as stretching or bending, but not cutting. A classic example is the number of holes in a manifold, which leads to the observation that a mug is topologically, but not geometrically, equivalent to a donut. In this sense, topology can be thought as more fundamental than geometry, since it does not rely on measures of length or angle \cite{nakahara2018geometry}.

\begin{figure*}[ht]
    \centering
    \begin{figframe}
    \includegraphics[width=\linewidth]{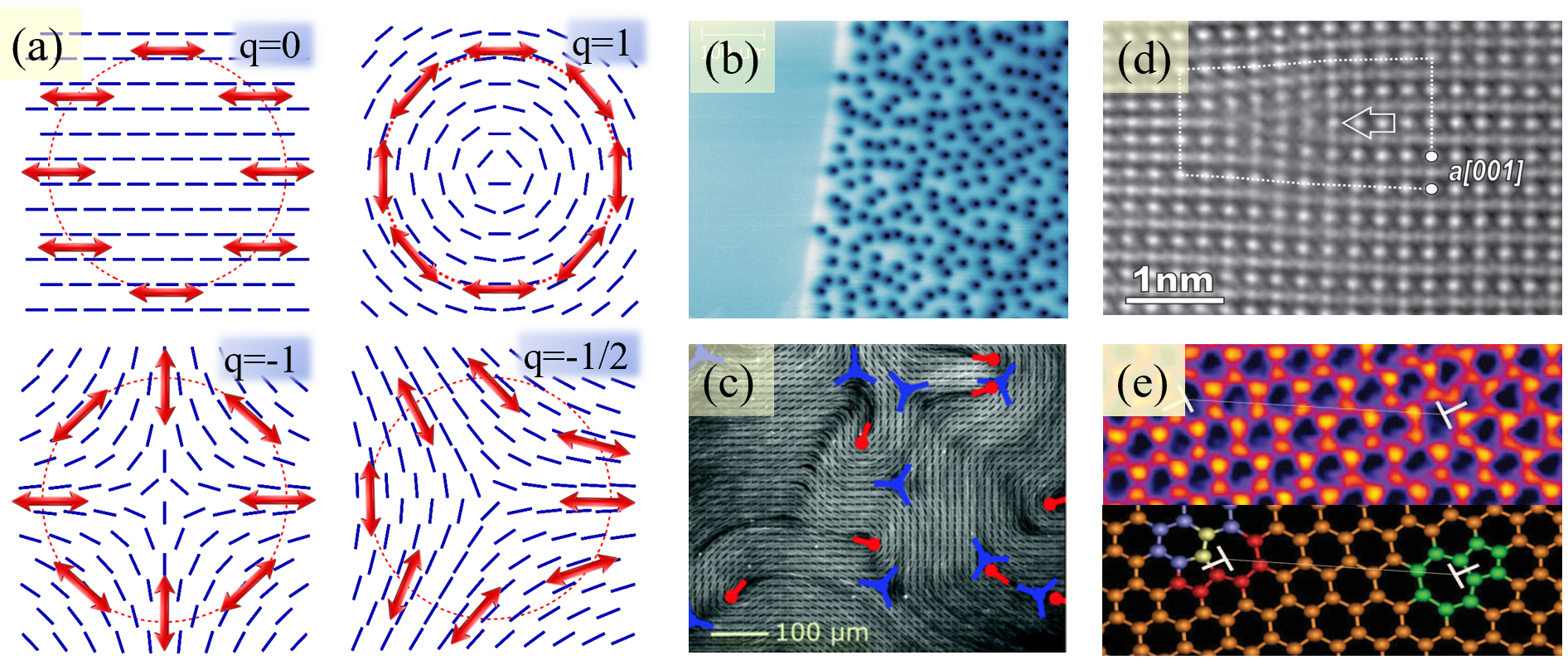}
     \end{figframe}
    \caption{\textbf{Topological defects.} \textbf{(a)}  Examples of topological defects in a director field with winding number $q=0,\pm 1,-1/2$. The dashed red circle indicates the loop used to compute the winding number $q$, while the red double-arrows emphasize the winding of the order parameter around the singularity. \textbf{(b)} Image of topological vortices in $200$ nm thick YBCO superconducting film taken by scanning SQUID microscopy at $4$ K. Figure taken from Ref.~\cite{Wells2015}. \textbf{(c)} Snapshot of a microtubule-based active nematic featuring multiple $q=+1/2$ disclinations (thick red dashes) and $q=-1/2$ disclinations (blue triads). Figure taken with permission from Ref.~\cite{PhysRevLett.127.197801}. \textbf{(d)} Electron microscopy image of a dislocation in SrTiO$_3$. The Burgers circuit yields a
 Burgers vector $a$[001], with $a$ the lattice spacing. The arrow indicates the direction of the dislocation line. Figure taken with permission from Ref.~\cite{PhysRevLett.95.225506}. \textbf{(e)} Top panel: High-resolution transmission electron microscopy image of a pair of edge dislocations (white symbols) in graphene. Bottom panel: Atomic model illustrating the $5$-$7$ pair structure of each dislocation. Figure taken with permission from Ref.~\cite{doi:10.1126/science.1217529}.
 }
     \label{fig:1}
\end{figure*}

The origin of topology is often traced back to Euler's solution of the Seven Bridges of K\"onigsberg problem (1736)~\cite{nash1997topology}, yet the field truly began to flourish in the 19th century through the pioneering contributions of Riemann, Poincar\'e, and others. From these early developments (Gauss, Poincar\'e, Lord Kelvin) to the modern era (Dirac, Aharonov-Bohm, Penrose), topology has become a central pillar of theoretical physics, with important consequences for applications.

In solid-state physics, topology, and in particular the notion of topological defects in elastic media, was already familiar to mathematicians such as Weingarten and Volterra in the early 20th century~\cite{weingarten1901surfaces,volterra1907annls}. Their work provided the foundations for the modern theoretical continuum frameworks later developed by Burgers~\cite{Burgers1980Dislocations} and Frank~\cite{Sluckin01091998}. However, the profound physical significance of TDs, particularly their role in controlling the mechanical response of crystals, emerged only later with the pioneering works of Frenkel (1926)~\cite{Frenkel1926}, Orowan, Polanyi, and Taylor (1934)~\cite{orowan1934kristallplastizitat,Polanyi1934,Taylor}. 

It was only much later that topology also rose to prominence in quantum condensed matter, driven by the discovery of the quantum Hall effect (1980)~\cite{10.1063/1.1611351,Hatsugai_1997} and topological insulators (2007)~\cite{RevModPhys.82.3045}. These breakthroughs demonstrated that, in some cases, topology can classify phases of matter and that topological defects may give rise to technologically relevant features, for example in quantum computation~\cite{doi:10.1126/science.1231473}. Based on these successes, wider applications that rely heavily on topological methods were developed from liquid crystals and biological systems to crystalline solids~\cite{nelson2002defects,kleinert1989gauge,Selinger:619845,Saw2017,Argento2025.04.14.648711}.

Across physics, the language of topology, particularly what in jargon is referred to as \textit{homotopy theory}, provides a powerful framework to identify, classify, and track defects in ordered systems. It is important to note that not all defects are topological~\cite{nakahara2003geometry,nelson2002defects}. Topological defects arise when an order-parameter field possesses a non-trivial homotopy group, meaning that it contains loops or surfaces that cannot be continuously contracted to a point. The defining feature of such topological defects is their protection: No local and continuous deformation of the field can remove them. This provides a rigorous mathematical basis for their robustness and persistence. Consequently, eliminating a topological defect requires either its migration out of the system via its boundary or its annihilation with a defect carrying the opposite topological charge.

To provide a concrete example (see \cite{Selinger:619845} for an extended discussion), let us consider two-dimensional nematic order. The order parameter is a unit director field $\hat{n}(x,y)$, with the identification $\hat{n}\equiv -\hat{n}$ reflecting the head--tail symmetry of each molecule. As a consequence, the order-parameter space is the real projective line, $\mathbb{RP}^1$, whose first homotopy group is non-trivial. This allows for the existence of topological defects characterized by a topological charge, or winding number,
\begin{equation}
q=\frac{1}{2\pi}\oint d\theta(x,y),\label{wind}
\end{equation}
where $\theta(x,y)$ denotes the local orientation angle of the director field and is defined modulo $\pi$, i.e., $\theta\equiv\theta+\pi$. Owing to this identification, the topological charge can take either integer or half-integer values. Physically, $q$ measures the total winding of the director orientation around a closed loop encircling the defect, with half-integer charges corresponding to a net rotation of an odd multiple of $\pi$. Four representative examples corresponding to different values of $q$ are shown in Fig.~\ref{fig:1}(a).

Beyond this textbook example, topological defects are ubiquitous, emerging in the swirl patterns of human hair~\cite{richeson2019euler}, in fingerprint structures~\cite{Fumeron2023}, and even as curvature-induced defects on soccer balls~\cite{kotschick_topology_2006}.
In physical systems, TDs manifest in many forms. Vortices in superconductors (Fig.~\ref{fig:1}(b)) and skyrmions in magnetic materials are iconic examples in quantum condensed matter. In soft matter, nematic and liquid-crystal textures naturally host defect lines and points (Fig.~\ref{fig:1}(c)). 

In crystalline solids, the most fundamental TDs are dislocations, which break translational symmetry and act as the microscopic ``carriers'' of plastic deformation. Arising from misaligned atomic planes, they can be directly imaged using electron microscopy. Some examples are illustrated in Fig.~\ref{fig:1}(d,e). Dislocations are a seminal concept in understanding crystal plasticity, as they are known to facilitate the rearrangement of the crystal lattice, relieving elastic stress, and dissipating energy. They play a crucial role in determining the yield point, flow properties, and fracture toughness of most crystalline metals, ceramics, and many polymers \cite{hull2011introduction,anderson2017theory}

Dislocations are characterized by a topological invariant known as the Burgers vector \cite{Burgers1980Dislocations}, defined as
\begin{equation}
\vec{b}=\oint d\vec{u},\label{burg}
\end{equation}
\noindent
where $\vec{u}=\vec{x}-\vec{x}_0$ is the displacement field, measuring the deviation of each particle position $\vec{x}$ from its corresponding site $\vec{x}_0$ in the ideal reference lattice. Geometrically, $\vec{b}$ quantifies the lattice mismatch accumulated along a closed loop encircling the dislocation core, corresponding to the lattice units required to close the loop, as illustrated in Fig.~\ref{fig:1}(d). 
The seminal works of Bilby and Kondo \cite{Bilby_1955,Kondo_1964} placed dislocations within the framework of differential geometry, formalizing their topological nature by identifying them as sources of torsion in an otherwise flat crystalline manifold.

Although in common parlance ``defect" carries a negative connotation, deviations from structural regularity in materials systems can give rise to important functionality. Materials scientists and engineers manipulate materials in large part through an understanding of how defects mediate a wide range of materials properties. As such, topological defects are far from mere mathematical curiosities: they encode essential physical information and are indispensable in describing numerous physical phenomena. For example, if dislocations are neglected, elasticity theory overestimates the maximum shear stress a crystal can sustain by almost four orders of magnitude~\cite{kleinert1989gauge}. Likewise, the presence, motion, and interactions of TDs are central to theoretical descriptions of two-dimensional melting~\cite{RevModPhys.89.040501}.

\begin{figure*}[ht!]
    \centering
    \begin{figframe}
    \includegraphics[width=\linewidth]{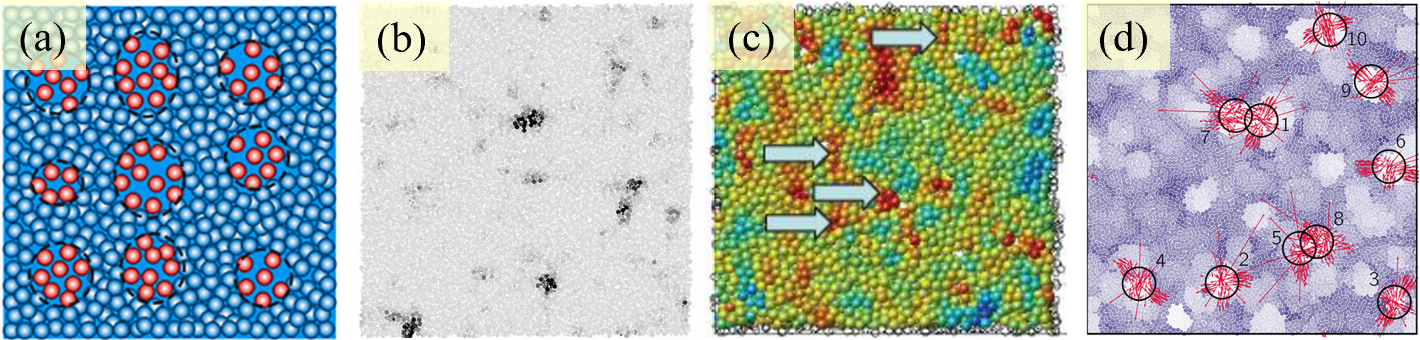}
     \end{figframe}
    \caption{\textbf{Defects in glasses.} \textbf{(a)} Shear transformation zones, localized regions undergoing plastic irreversible rearrangements. Figure taken with permission from \cite{CHEN2022107496}. \textbf{(b)} Plastic events are traditionally identified using the non-affine descriptor $D^2_{\mathrm{min}}$~\cite{falk1998dynamics}, that exhibits large values (black color) in localized regions (``soft spots''). \textbf{(c)} Experimental identification of structural rearrangements and shear transformation zones in colloidal glasses~\cite{doi:10.1126/science.1149308}. \textbf{(d)} Localized plastic defects with quadrupolar structure in a simulated glass~\cite{PhysRevLett.126.015501}.}
    \label{fig:2b}
\end{figure*}

\section*{Glasses and amorphous solids}
Humankind has produced glasses for thousands of years by cooling glass-forming liquids, but it was only in the 20th century that X-ray studies revealed similarities between the static structure factor of glasses and liquids, providing the first evidence for their common atomic scale structural disorder. Oxide glasses, metallic glasses, polymers, gels, granular materials, and biological systems often derive their unique properties from their underlying structural disorder, which allows for material structures that continuously adapt to variations in composition and thermomechanical history~\cite{varshneya2019fundamentals}.

Despite their ubiquity, compared to their crystalline cousins, we are less able to relate amorphous structures to properties. Some properties of glasses can be connected to their short range order~\cite{binder2011glassy,varshneya2019fundamentals}, but others, such as their elastic properties, depend on medium range order (MRO) on length scales of 4-10 particle diameters. At present we lack deep insight into the nature of the MRO since it has only recently become possible to experimentally probe the structural order on these scales \cite{Yang_2021,Lichtenstein2012}.

Theoretical descriptions of structure often neglect MRO since the typical quantities used in standard liquid state theory, like the static structure factor, do not adequately capture features on these length scales. Ongoing work seeks to elucidate the relationships between local structure, which is often determined by interactions between particles due to compositional variation, see for example~\cite{ROYALL20151}, with structures on larger length scales. Once the structural features of the glass-forming liquid are identified, one needs to relate them to the evolution and responses of the system, a formidable independent problem~\cite{varshneya2019fundamentals,binder2011glassy}. 

We note that while understanding the structure is certainly important to understand the mechanical properties of glasses, it may also be highly relevant to the mechanisms responsible for the dramatic slowing down of the glass-former upon cooling~\cite{binder2011glassy,CAVAGNA200951}.

\section*{Defects in glasses \& challenges from disorder}
The notion of defects in glasses is not new. Doping, e.g., the introduction of metallic atoms into an oxide glass matrix, can give rise to color. Surface defects present in a glass sample are highly relevant to their fracture properties. Such structural modifications have long been studied and exploited~\cite{varshneya2019fundamentals}. Given the success of dislocation theory in applying topological concepts in the crystalline context, here we focus on recent developments and open questions relating topological defects to the mechanical responses of amorphous solids.

A first notion of topology arises in bonding and neighbour networks, such as continuous random networks and network-forming glasses \cite{phillips1985constraint,egami2007glass,gupta2007topological,mauro2011topological}. Here, topology refers to the connectivity of a discrete network rather than to a continuous order-parameter field, and defects can be identified directly from changes in this connectivity, for example through Voronoi/Delaunay tessellations \cite{PhysRevMaterials.4.125601}. In two dimensions, the coordination charge $q_i=6-z_i$ is integer valued and obeys the global constraint $\sum_i q_i=0$ for a flat system with periodic boundaries, naturally identifying five- and seven-fold coordinated particles as oppositely charged objects. In crystals these coincide with conventional disclinations, whereas in glasses their quantization follows from the discrete neighbour network itself, rather than from the homotopy of a broken-symmetry order parameter. While for simple glass-forming systems such a bond counting allows to obtain interesting insights, it looses its power in the case of systems with strong polydispersity, and hence other approaches are needed.

The structural disorder of glasses and the resulting lack of an obvious order parameter might lead to the conclusion that the formal concept of topological defect is not useful for such materials. This perhaps explains why many attempts to identify defects in amorphous systems have not explicitly employed topological measures but focused on local atomic arrangements, e.g., Refs.~\cite{Srolovitz01101981,rivier1982topological,sachdev1985order,sadoc1987hierarchy,  lee2011networked, ding2014full,Lieou2023}. 

Attempts in the 1970's and 80's to extend topological concepts from crystals to amorphous materials failed to result in a systematic approach to defining topological defects in amorphous solids. Gilman advocated that glasses contain dislocations with variable Burgers vectors that leave debris in their wake \cite{gilman1973flow,gilman1975mechanical}. Subsequent early simulation work by Chaudhari, Levi, and Steinhardt \cite{chaudhari1979edge} observed that dislocations could be produced by applying a Volterra construction \cite{volterra1907annls}. In some cases these were observed to be stable while in other cases they were found to decompose into disclinations \cite{steinhardt1981point}. The final verdict of these studies, as well as their assumption that the local density is the appropriate order parameter for glasses, remains open and subject to debate. Nelson proposed a geometrical-frustration picture in which locally preferred icosahedral order is naturally accommodated in curved space, while its realization in Euclidean three-dimensional space requires a network of topological disclination lines \cite{PhysRevLett.50.982,PhysRevB.28.5515}. While conceptually appealing, the extent to which this construction provides a direct description of defects in generic glasses remains unclear.

Later investigations of plasticity in computer models of amorphous solids prompted researchers to reconsider topological measures. Inspired by Argon's early observations of rearrangements in sheared bubble rafts \cite{argon1979plastic}, Falk and Langer proposed a picture of persistent localized defects termed ``shear transformation zones" or STZs \cite{falk1998dynamics} (see Fig.~\ref{fig:2b}(a-c)). Subsequent investigations showed that the resulting rearrangements result in quadrupolar elastic fields, as one would expect from a sheared Eshelby inclusion in an elastic medium, and that these signatures are also present in the vibrational modes of the system, e.g. Fig.~\ref{fig:2b}(d) and Refs.~\cite{manning2011vibrational,maloney2006amorphous,tanguy2010vibrational,rottler2014predicting,tanguy2015vibration}. An analysis of multiple signatures correlated with plastic activity in simulated amorphous systems \cite{PhysRevMaterials.4.113609} that compared purely structural \cite{tong2018revealing,kawasaki2007correlation,rieser2016divergence,piaggi2017entropy}, machine-learned \cite{cubuk2015identifying,schoenholz2016structural,ronhovde2011detecting,boattini2019unsupervised}, isotropic \cite{tanguy2010vibrational,manning2011vibrational,widmer2008irreversible,zylberg2017local,tong2014order} and anisotropic linear response \cite{tsamados2009local,schwartzman2019anisotropic,xu2021atomic}, and nonlinear shear response measures \cite{lerner2016micromechanics,xu2017strain,xu2018predicting} showed that, while all exhibit some predictive capacity, the initial plastic response of high-quench-rate glass structures are best predicted by linear measures of the nonaffine response. The same study showed that ultra-stable glasses that sustain larger shears before deforming are only well-predicted by non-perturbative measures \cite{patinet2016connecting,Barbot2018LocalSolids}. This analysis left open the question of whether any of these measures relates to the topology of underlying, as yet unspecified, order parameter fields.

\section*{Application of topological methods to disordered systems}
Recent investigations have sought to identify topological defects in amorphous solids and point the way toward their formal definition in terms of an underlying order parameter field. A universally accepted order parameter has yet to emerge, and the methods used to quantify TDs has varied across studies.

Perhaps the simplest definition consists in considering the local orientation of a 2D vector field $\theta(x,y)$ and computing the associated winding number $q$, as defined in Eq.~\eqref{wind}. When this construction is applied to a vector field (rather than to a director field, as in Fig.~\ref{fig:1}(a)), the elementary defects are vortices ($q=+1$) and anti-vortices ($q=-1$). Notably, the negatively charged defects exhibit a quadrupolar symmetry, analogous to that clearly visible in the director field shown in Fig.~\ref{fig:1}(a).

\begin{figure*}[htbp]
\centering
\begin{figframe}
\begin{minipage}{0.22\textwidth}
  \includegraphics[width=\textwidth]{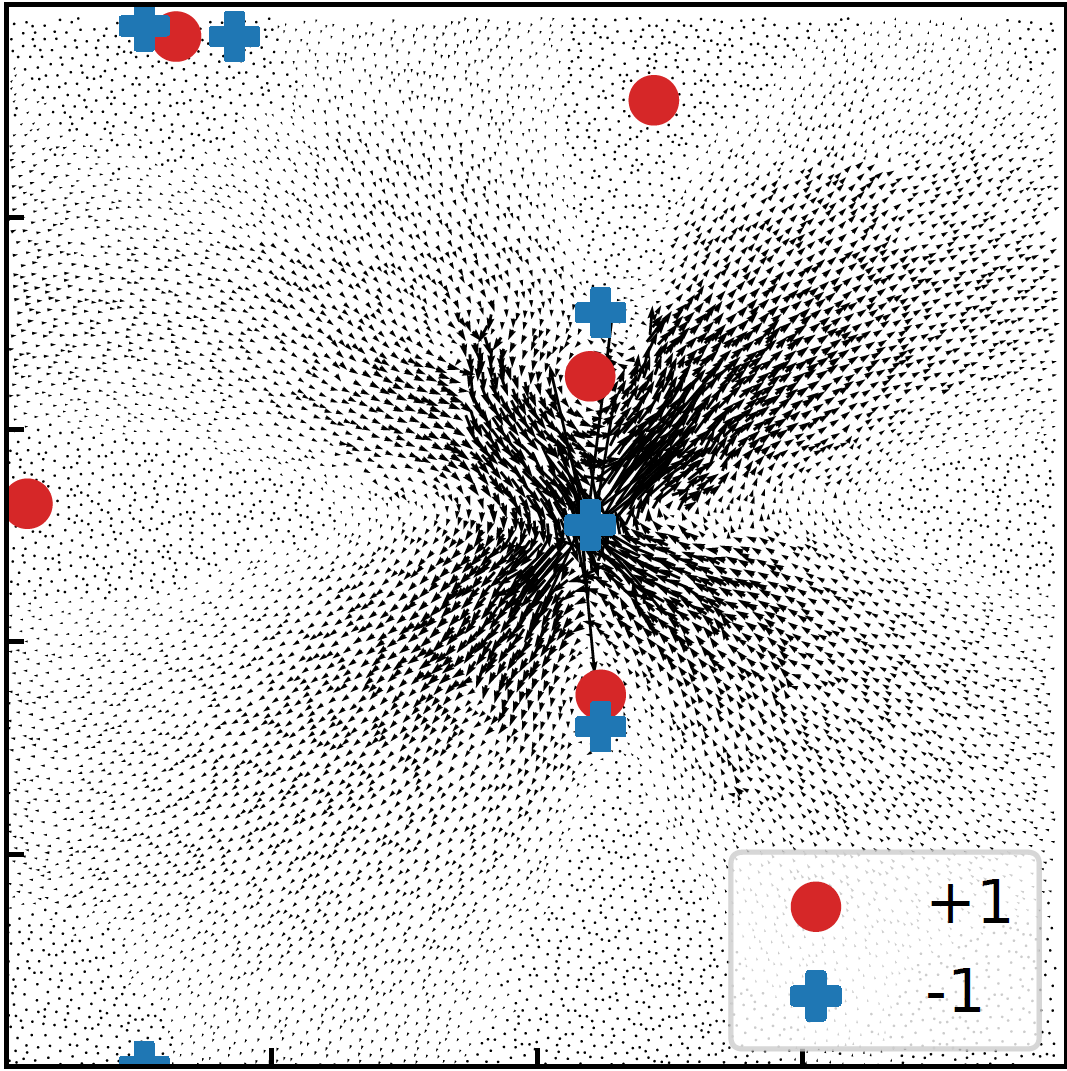}
\end{minipage}
\hspace{0.01\textwidth}
\begin{minipage}{0.23\textwidth}
  \vtop{\footnotesize
  \textbf{(a)} Non-affine displacement field during a plastic event in a simulated 2D binary Lennard-Jones glass. A quadrupolar Eshelby-like event is visible in the center of the box. \reddisk{} and \includegraphics[height=0.185cm]{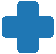} symbols indicate, respectively, the positions of the $\pm 1$ topological defects. Figure adapted from \cite{PhysRevE.109.L053002}.
  }
\end{minipage}
\qquad
\begin{minipage}{0.23\textwidth}
  \includegraphics[width=\textwidth]{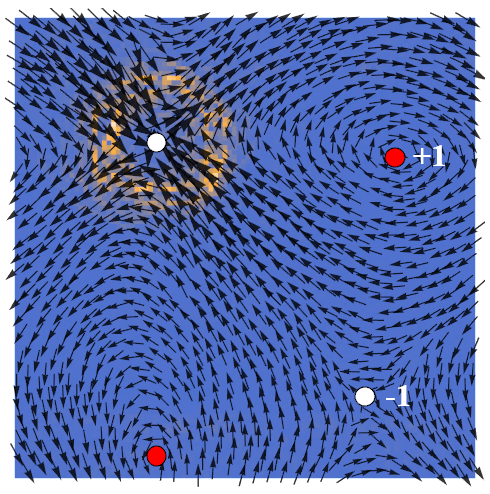}
\end{minipage}
\hspace{0.01\textwidth}
\begin{minipage}{0.23\textwidth}
  \vtop{\footnotesize
  \textbf{(b)} Displacement field for a simulated 2D Lennard-Jones glass. \includegraphics[height=0.29cm]{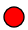} and \includegraphics[height=0.27cm]{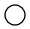} symbols indicate, respectively, the positions of the $\pm 1$ topological defects. The background color represents the magnitude of the local Burgers vector. A Burgers ring surrounding the quadrupolar Eshelby-like event is evident. Figure adapted from \cite{bera2025burgers}.}
\end{minipage}

\vspace{0.2cm}

\begin{minipage}{0.23\textwidth}
  \includegraphics[width=\textwidth]{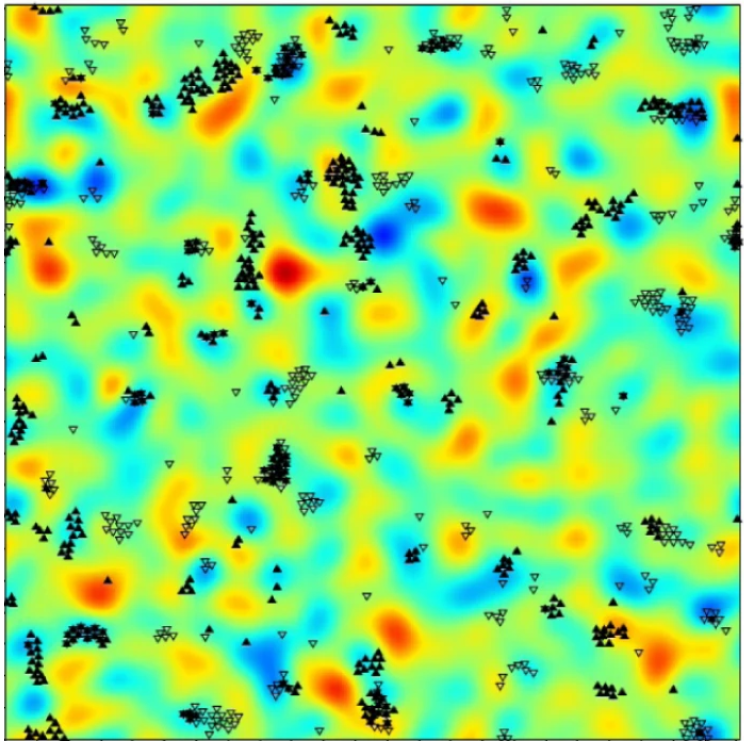}
\end{minipage}
\begin{minipage}{0.23\textwidth}
  \vtop{\footnotesize
  \textbf{(c)} Correlation between negative TDs in the eigenvector fields and plastic events under shear deformation in  a simulated 2D glass. The background color is the charge density of TDs from negative values (\color{blue}blue\color{black}) to positive ones (\color{red}red\color{black}). Black symbols are plastic events from a shear in the positive and negative $x$ direction. Figure adapted from \cite{wu2023topology}.}
\end{minipage}
\qquad
\begin{minipage}{0.25\textwidth}
  \includegraphics[width=\textwidth]{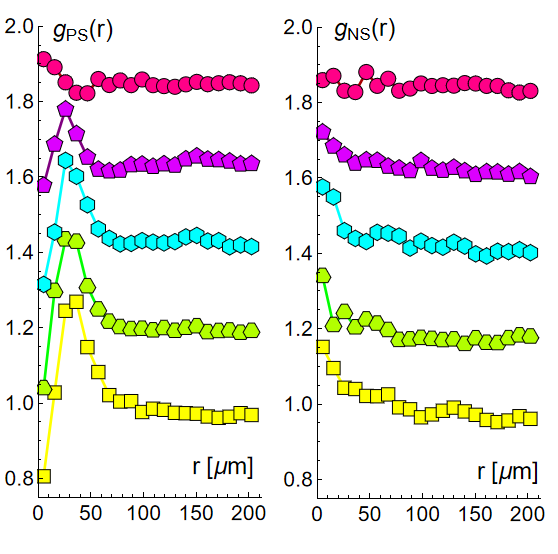}
\end{minipage}
\begin{minipage}{0.23\textwidth}
  \vtop{\footnotesize
  \textbf{(d)} Radial pair correlation function between positive (P) and negative (N) TDs in the eigenvector fields and soft spots (S) under shear deformation in an experimental 2D colloidal glass for increasing frequencies, shifted vertically, yellow being the lowest. Figure adapted from \cite{Vaibhav2025}.}
\end{minipage}

\vspace{0.2cm}

\begin{minipage}{0.23\textwidth}
  \includegraphics[width=\textwidth]{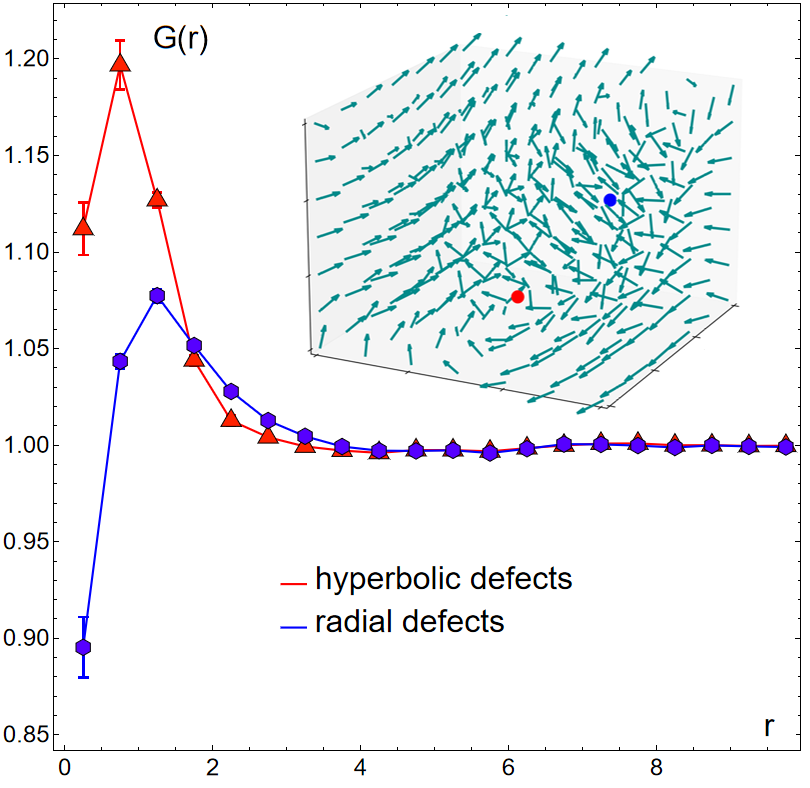}
\end{minipage}
\begin{minipage}{0.23\textwidth}
  \vtop{\footnotesize
  \textbf{(e)} Average correlation function $G(r)$ between soft spots and hyperbolic (H) and radial (R) hedgehog defects in the eigenvector field in a simulated 3D glass. The inset shows a snapshot of an eigenvector field containing a positive (\reddisk{}) and negative (\bluedisk{}) TD. Figure adapted from \cite{bera2025hedgehogtopologicaldefects3d}.}
\end{minipage}
\qquad
\begin{minipage}{0.25\textwidth}
  \includegraphics[width=\textwidth]{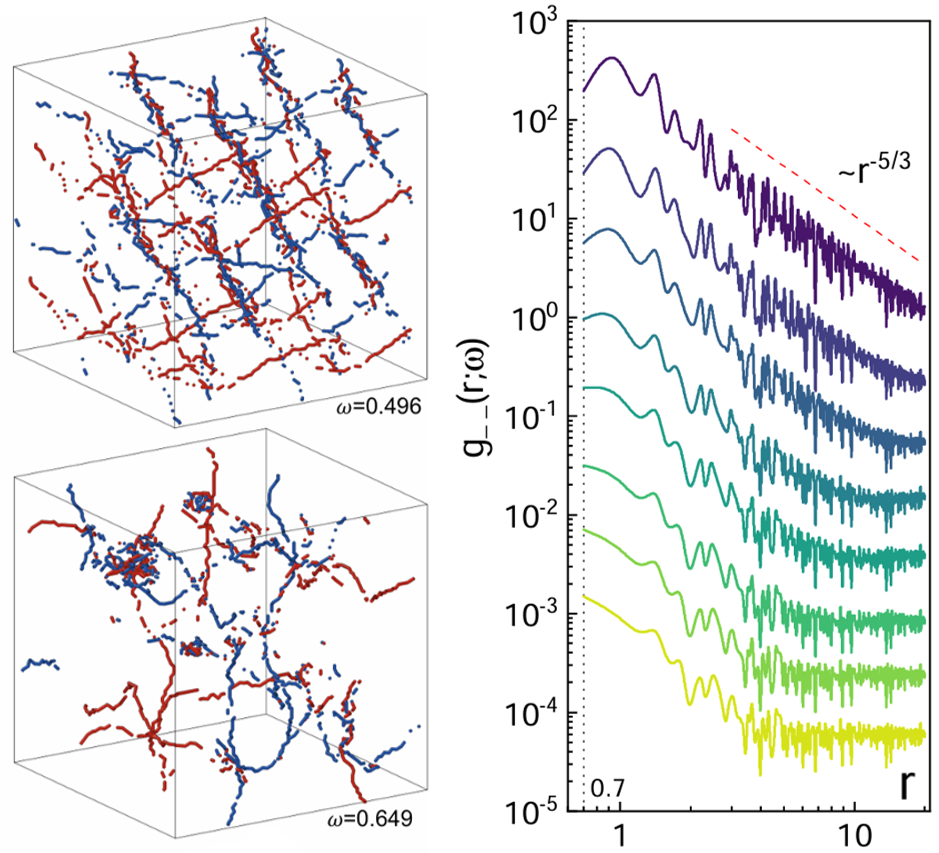}
\end{minipage}
\begin{minipage}{0.23\textwidth}
  \vtop{\footnotesize
  \textbf{(f)} Vortex-lines in two low-frequency eigenvector fields obtained from a simulated 3D glass. \color{red}Red \color{black} and \color{blue}blue \color{black}correspond respectively to positive and negative defects. The vertical panel presents the radial pair correlation between anti-vortex lines from low (dark blue) to low (yellow) frequencies. Figure adapted from \cite{wu2024geometrytopologicaldefectsglasses}.}
\end{minipage}

\vspace{0.2cm}

\begin{minipage}{0.25\textwidth}
  \includegraphics[width=\textwidth]{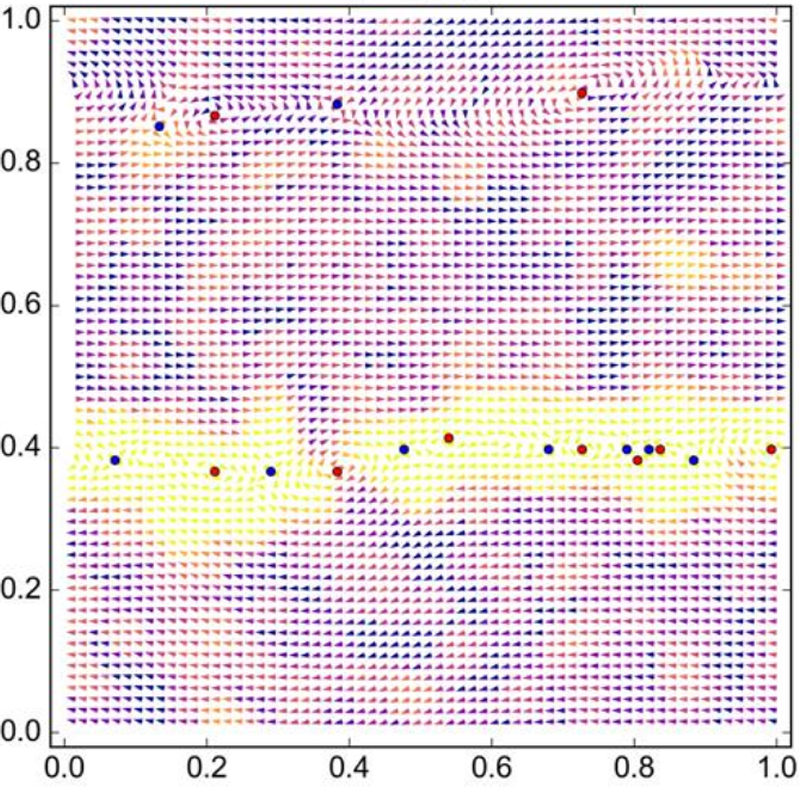}
\end{minipage}
\begin{minipage}{0.23\textwidth}
  \vtop{\footnotesize
  \textbf{(g)} Displacement field of a 2D simulated glass under shear deformation. The background color indicates the amplitude of the local non-affine measure $D^2_{\text{min}}$ showing a horizontal shear band. \reddisk{} and \bluedisk{} are respectively positive/negative TDs. Figure adapted from \cite{bera2025microscopic}.}
\end{minipage}
\qquad
\begin{minipage}{0.23\textwidth}
  \includegraphics[width=\textwidth]{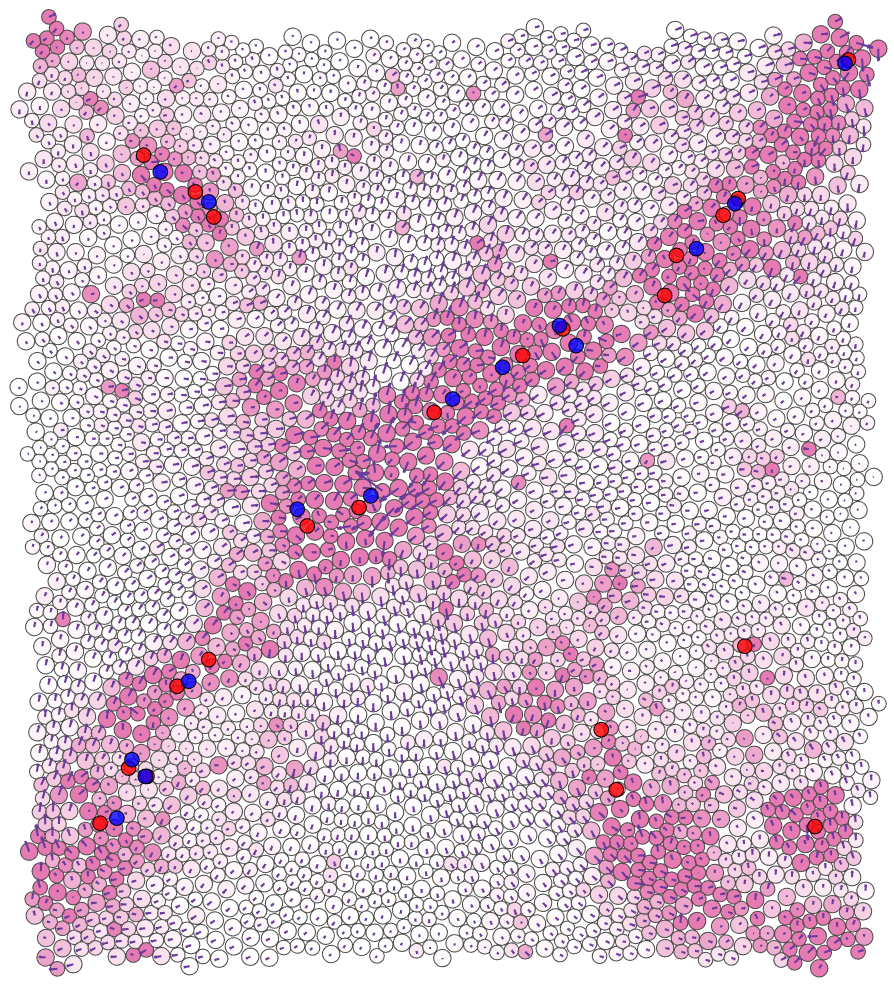}
\end{minipage}
\begin{minipage}{0.23\textwidth}
  \vtop{\footnotesize
  \textbf{(h)} Particle-level displacement field in a 2D granular glass under simple shear. The background color indicates the magnitude of the local non-affine measure $D^2_{\text{min}}$ showing a horizontal shear band. \reddisk{} and \bluedisk{} are respectively positive/negative TDs. Figure adapted from \cite{wang2025topological}.}
\end{minipage}

\vspace{0.2cm}

\begin{minipage}{0.24\textwidth}
  \includegraphics[width=\textwidth]{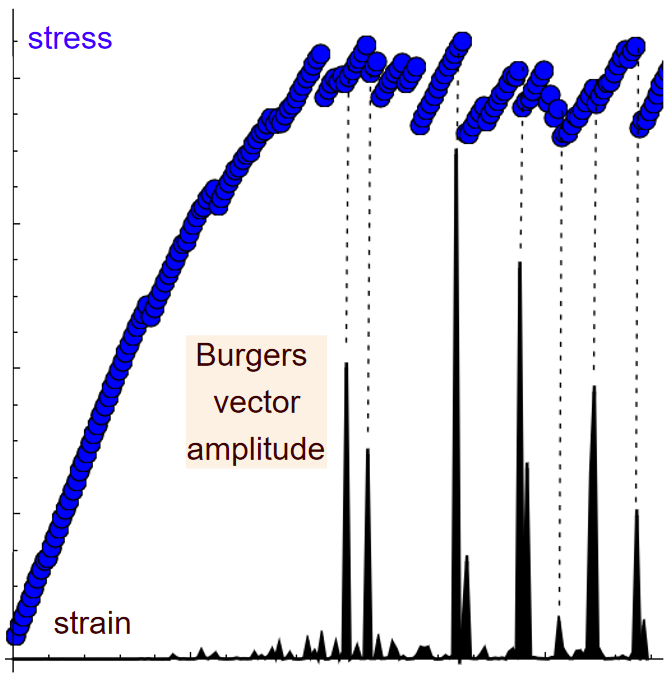}
\end{minipage}
\begin{minipage}{0.22\textwidth}
  \vtop{\footnotesize
  \textbf{(i)} The \color{blue}blue \color{black} data show the stress-strain curve of a 2D simulated glass under shear deformation. The black spikes represent the norm of the Burgers vector that coincide with the plastic stress drops (vertical dashed lines). Figure adapted from \cite{PhysRevLett.127.015501}.}
\end{minipage}
\qquad
\begin{minipage}{0.25\textwidth}
  \includegraphics[width=\textwidth]{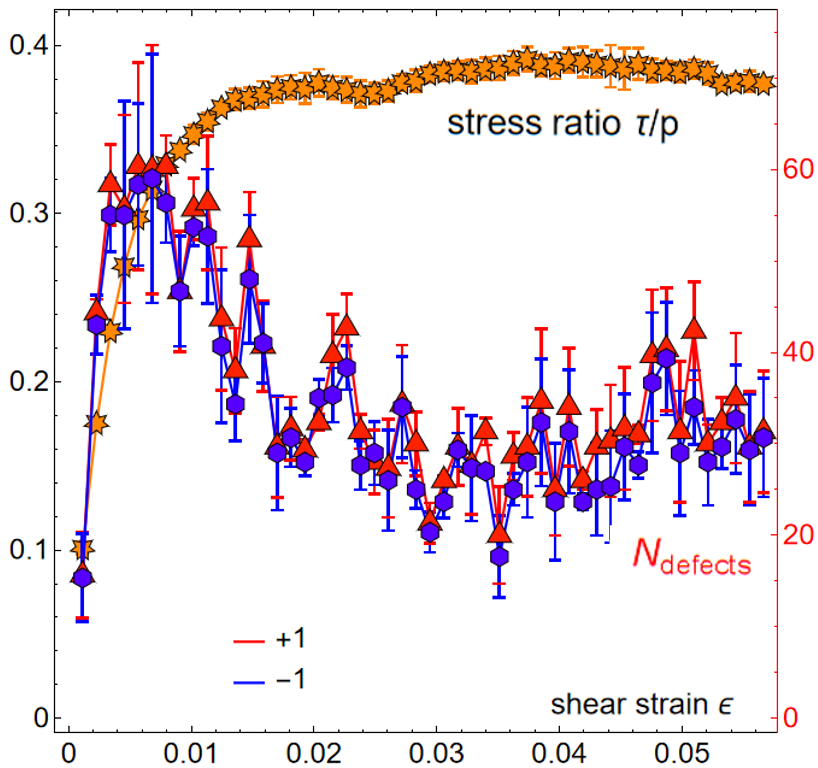}
\end{minipage}
\begin{minipage}{0.23\textwidth}
  \vtop{\footnotesize
  \textbf{(j)} Stress-strain curve (\color{orange} orange \color{black} symbols) for a 2D amorphous granular packing under simple shear. \color{Red}Red \color{black} and \color{blue}blue \color{black} symbols represent the total number of $\pm 1$ TDs in the displacement vector field at each strain step. Figure adapted from \cite{wang2025topological}.}
\end{minipage}
\end{figframe}
\vspace{-14pt}
\caption{\textbf{Overview of representative results from simulations and experiments regarding topological defects in amorphous solids.}}
\label{fig:fullpanel}
\end{figure*}

Applying this procedure to the non-affine irreversible displacement field $\vec{u}$ of a glass under mechanical deformation reveals interesting features in the spatial distribution of $\pm 1$ TDs~\cite{PhysRevE.109.L053002}. Panel (a) of Fig.~\ref{fig:fullpanel} shows the particle displacement field following a plastic event in a simulated two-dimensional binary Lennard--Jones (LJ) glass under shear. The non-affine field displays a pronounced quadrupolar, Eshelby-like structure. Strikingly, the isolated $q=-1$ defect coincides almost perfectly with the core of this quadrupolar pattern.
Such vortex-like defects in the dynamical displacement field also appear to play a significant role in active granular systems. In particular, they exhibit a strong correlation with the locations of activity-induced structural rearrangements \cite{doi:10.1073/pnas.2510873123}. 

At the same time, several additional TDs appear in the non-affine displacement field away from the main plastic event in Fig.~\ref{fig:fullpanel}(a). These secondary defects are not spurious and contribute to the full displacement field generated by the event. The results therefore suggest an intriguing correspondence between Eshelby-like quadrupolar structures and anti-vortices, a connection that is consistent with the symmetry they share.
Figure~\ref{fig:fullpanel}(a) also makes clear that identifying vortex-like TDs in the displacement field alone does not unambiguously pinpoint the location of the dominant plastic event, where irreversible rearrangements are concentrated. In fact, one could reasonably argue that the correspondence is not one-to-one (see also Supplementary Information in \cite{Hu2025}).
This arises from the definition of the winding number which depends only on the local orientation of the vector field and is completely insensitive to its amplitude. Mathematically, the winding number detects singularities of the phase and does not directly encode information about the magnitude of the underlying displacement. The latter has been shown to be crucial for identifying the physically dominant plastic rearrangement and connecting to the resulting elastic Eshelby field \cite{PhysRevE.109.L053002}. 

A possible way to overcome this limitation is to adopt a definition of TD that is sensitive to the amplitude of the vector field. One proposal, put forward in \cite{PhysRevLett.127.015501,Landry}, relies on the concept of a continuous Burgers vector reviewed by Kl\'eman \cite{RevModPhys.80.61}. The idea is to apply the standard definition of the Burgers vector, Eq.~\eqref{burg}, directly to the non-affine displacement field of a glass undergoing mechanical deformation. It is important to emphasize that, due to the intrinsically continuous nature of the resulting Burgers vector, this object is not topological in the strict sense.
Figure~\ref{fig:fullpanel}(b) shows a representative example of this construction for a two-dimensional Lennard--Jones glass. An Eshelby-like plastic event is visible in the top-left corner. As in Fig.~\ref{fig:fullpanel}(a) we note that not all $-1$ TDs are equally relevant for the generation of plastic events since either the magnitude of the field is too small, or the field is shielded by an adjacent $+1$TD. 
In contrast, the background color encodes the magnitude of the continuous Burgers vector, $|\vec{b}|$, which exhibits a pronounced, ring-shaped peak surrounding the main Eshelby-like event~\cite{bera2025burgers}. This signal clearly identifies the region where the displacement amplitude is largest. The result indicates that the amplitude of the vector field carries essential physical information and cannot be neglected. If one insists on characterizing plasticity solely in terms of phase-defined TDs, an additional weighting procedure that incorporates amplitude information becomes necessary. This is also evident in Ref.~\cite{PhysRevE.109.L053002} where the magnitude of the displacements that give rise to the $-1$ TDs are shown to correlate with the strength of the resulting Eshelby elastic field.
Interestingly, the concept of the Burgers vector has already been extended to experimental diffraction measurements \cite{Liu:ib5129}, highlighting its potential correlation with local structural transformations and with structural indicators such as the degree of centrosymmetry. 

We also note that a complementary proposal for identifying defects in eigenvector fields has recently been introduced in Ref.~\cite{huang2025geometric}. This work proposes a geometric quantity, closely related to the Nye dislocation density \cite{NYE1953153}, which characterizes the density of dislocations in crystals, as an alternative candidate for predicting plastic events from the normal modes of the undeformed configuration. Similar geometric densities, which reduce in crystals to standard predictors of plasticity such as dislocation and disclination densities, accurately identify plastic soft spots characterized by large $D^2_{\mathrm{min}}$ values in two-dimensional amorphous solids \cite{Wang2026UnifiedGeometricPlasticity}.

Up to this point, we have discussed TDs in the displacement field during deformation. In this sense, they provide a novel topological characterization of irreversible rearrangements in glasses. However, this description remains essentially retrospective and lacks predictive power. A natural question therefore arises: can one predict the location of plastic events directly from the initial, undeformed configuration?

Previous works have shown that this is indeed possible by analyzing the properties of vibrational modes around the initial state \cite{10.1063/1.3265983,tanguy2010vibrational,manning2011vibrational,PhysRevLett.107.108301,rottler2014predicting}, even in active systems \cite{hain2026usingforcelandscapeactive}. However, most of these approaches rely either on phenomenological criteria or on machine-learning techniques, which may obscure the underlying physical mechanism. In contrast, Ref.~\cite{wu2023topology} (see also \cite{Baggioli2023}) proposed a conceptually different route. The idea is to apply the notion of TDs, specifically, vortex-like defects defined via the winding number $q$ in Eq.~\eqref{wind}, to the eigenvector fields at a given frequency $\omega$, obtained from a normal mode analysis of the configuration prior to deformation.

An average over $\omega$, restricted to the low-frequency sector, makes it possible to identify regions in the sample with a high density of negative TDs. As shown in Fig.~\ref{fig:fullpanel}(c), these regions exhibit a strong correlation with the locations of plastic events in the sheared sample. This finding suggests that so-called soft spots are associated with an enhanced density of $q=-1$ TDs, which, owing to their saddle-like character, correspond to mechanically unstable regions of the system.

In general, eigenvector fields are difficult to access experimentally, as their determination requires either an extremely accurate knowledge of the interparticle interactions or the construction of correlation matrices based on real-time particle tracking (as in colloidal suspensions, granular systems, or complex plasmas). Remarkably, Ref.~\cite{Vaibhav2025} succeeded in accomplishing this task in a two-dimensional binary colloidal mixture with magnetic interactions.
 This study confirmed the simulation results of Ref.~\cite{wu2023topology}, demonstrating a clear short-range correlation between $q=-1$ defects and plastic spots (reproduced in Fig.~\ref{fig:fullpanel}(d)), and thereby provides solid experimental support for this emerging framework.

Interestingly, this approach has recently been extended beyond amorphous solids, notably to dodecagonal quasicrystals \cite{bedollamontiel2025strikingsimilaritiesdynamicsvibrations} and conventional crystals \cite{HUANG2025106274}. In particular, Ref.~\cite{HUANG2025106274} reported compelling evidence for a connection between vortex-like structures in the eigenvector fields of low-frequency modes and well-established structural defects in crystals, such as dislocations. These findings open the door to applying this emerging topological framework to ordered solids and suggest a possible unifying perspective across disordered and crystalline systems.

When extending the analysis to three-dimensional systems, the landscape of topological defects (TDs) becomes considerably richer, as both point-like and line-like defects can be defined. A first approach adopted in \cite{10.1093/pnasnexus/pgae315}, was to approximate the three-dimensional system as a stack of two-dimensional slices, to which the definition of vortex-like defects introduced in \cite{wu2023topology} could be directly applied. This procedure led to interesting observations, in particular the clustering of vortex-like defects preceding plastic events and their subsequent annihilation afterwards. However, from a formal standpoint, this construction is not entirely satisfactory, as it does not treat the system as a genuinely three-dimensional vector field and therefore cannot fully capture the intrinsic topology of defects in 3D.

In contrast, Ref.~\cite{bera2025hedgehogtopologicaldefects3d} focused on topological point defects, commonly referred to as hedgehog defects, identified in the three-dimensional eigenvector field of a simulated glass. Within this framework, it became clear that topology alone is insufficient to characterize plastic events, as it does not encode information about the geometric structure of the associated defects.

When geometric features are taken into account, a more refined picture emerges: so-called hyperbolic defects (as opposed to radial ones) exhibit a strong correlation with plastic rearrangements under deformation (see Fig.~\ref{fig:fullpanel}(e)). This finding is, in some sense, expected. Hyperbolic defects share the same saddle-point structure as anti-vortices in two dimensions, emphasizing that these unstable configurations are the key ingredient linking defect structure to irreversible plastic events.

An alternative three-dimensional approach, based on vortex line defects, was introduced in \cite{wu2024geometrytopologicaldefectsglasses}. Within this framework, $-1$ vortex lines characterized by a saddle-point structure again display a strong correlation with plastic rearrangements. Even more intriguingly, the spatial organization of these topological defects exhibits the same scale-invariant power-law behavior which in that study was observed for plastic events identified through $D^2_{\mathrm{min}}$ clusters (see  Fig.~\ref{fig:fullpanel}(f)). This striking similarity confirms the deep connection between topological defects and plastic activity found in two-dimensional systems and thus calls for further investigations.

In many systems, including metallic glasses and granular materials, sufficiently large applied strains lead to the localization of plastic deformation within narrow regions known as \textit{shear bands}. It has recently been proposed \cite{PhysRevB.110.014107}, and subsequently confirmed in numerical simulations of two-dimensional systems \cite{bera2025microscopic}, that vortex-like defects organize along these shear bands, forming chains of alternating $\pm 1$ topological charges (see Fig.~\ref{fig:fullpanel}(g)). This observation is consistent with previous literature \cite{PhysRevLett.109.255502,kumar2026analytic,doi:10.1073/pnas.1506531112,doi:10.1073/pnas.2427082122,PhysRevLett.119.195503,Bian2020,PhysRevLett.128.245501,PhysRevLett.123.195502,SOPU2021100958,https://doi.org/10.1002/adma.202212086,TORDESILLAS2016215}.

Recently, Ref.~\cite{wang2025topological} reported an experimental study of a two-dimensional granular system subjected to shear deformation, elucidating the evolution of vortex-like defects as the applied shear strain increases. The topological defects are observed to migrate toward regions where shear strain localizes and plastic rearrangements occur, exhibiting a strong correlation with variations in the sixfold bond-orientational order parameter. Furthermore, their spatial distribution evolves from an approximately isotropic configuration to a markedly anisotropic one, with the majority of defects aligning along the developing shear bands (see Fig.~\ref{fig:fullpanel}(h)). A clear causal connection between these observations, as well as a unifying theoretical framework to explain them, remains to be established.

In \cite{PhysRevLett.127.015501}, a strong correlation was identified between global stress drops in the stress–strain curve of a simulated two-dimensional glass and the average amplitude of the Burgers vector (see Fig.~\ref{fig:fullpanel}(i)). This finding supports the idea that the latter can serve as both a probe and a predictor of the yielding instability, and global plastic events in general. These results were subsequently confirmed in other systems \cite{10.1093/pnasnexus/pgae315,Liu:ib5129}, further reinforcing the proposed connection between defect statistics and macroscopic mechanical failure.

On the other hand, \cite{wang2025topological} investigated vortex-like topological defects in the displacement field of an experimental two-dimensional granular system subjected to shear deformation. Remarkably, as shown Fig.~\ref{fig:fullpanel}(j), the number of these defects exhibits a pronounced peak as the system approaches the plastic flow regime, followed by a rapid decrease toward a steady state. This steady state is characterized by oscillatory cycles of defect annihilation and formation of defect clusters, closely resembling the behavior reported in \cite{10.1093/pnasnexus/pgae315}.

Finally, it is worth noting that Wang and collaborators~\cite{Cao2018} have examined the role of defects in plasticity within three-dimensional experimental granular systems. Their findings suggest that highly distorted coplanar tetrahedra act as the structural defects responsible for microscopic plastic rearrangements in disordered granular packings. Whether these entities can be rigorously classified as topological defects in a formal sense, and how they relate to the concepts discussed in this \textit{Perspective}, remains an open question.

\begin{table*}[t]
\centering
\scriptsize
\renewcommand{\arraystretch}{1.30}
\setlength{\tabcolsep}{3.5pt}

\newcommand{\TAr}[1]{\parbox[t]{0.090\textwidth}{\raggedright #1}}
\newcommand{\TAc}[1]{\parbox[t]{0.105\textwidth}{\centering #1}}
\newcommand{\TBc}[1]{\parbox[t]{0.120\textwidth}{\centering #1}}
\newcommand{\TCl}[1]{\parbox[t]{0.150\textwidth}{\raggedright #1}}

{\color{black}
\begin{tabular}{@{}cccccccc@{}}
\toprule

\TAr{\centering\textbf{Defect}\\\textbf{type}} &
\TAc{\textbf{Main}\\\textbf{reference}} &
\TAc{\textbf{Underlying}\\\textbf{field}} &
\TBc{\textbf{Spatial}\\\textbf{dimensionality}} &
\TBc{\textbf{Defect}\\\textbf{dimensionality}} &
\TCl{\centering\textbf{Topological or}\\\textbf{geometric charge}} &
\TAc{\textbf{Experimental}\\\textbf{evidence}} &
\TAc{\textbf{Role}} \\[0.2cm]

\midrule

\rowcolor{gray!12}
\multicolumn{8}{@{}l}{\strut\textbf{Discrete topological defects}} \\

\TAr{Vortex} &
\TAc{Ref.~\cite{wu2023topology}} &
\TAc{Eigenvector\\field} &
\TBc{$2$D} &
\TBc{$0$D (point)} &
\TCl{Winding number, Eq.~\eqref{wind}} &
\TAc{Ref.~\cite{Vaibhav2025}} &
\TAc{Predictor} \\

\addlinespace[2pt]

\TAr{Vortex line} &
\TAc{Ref.~\cite{wu2024geometrytopologicaldefectsglasses}} &
\TAc{Eigenvector\\field} &
\TBc{$3$D} &
\TBc{$1$D (line)} &
\TCl{Winding number, Eq.~\eqref{wind}} &
\TAc{---} &
\TAc{Predictor} \\

\addlinespace[2pt]

\TAr{Vortex} &
\TAc{Ref.~\cite{PhysRevE.109.L053002}} &
\TAc{Displacement\\vector field} &
\TBc{$2$D} &
\TBc{$0$D (point)} &
\TCl{Winding number, Eq.~\eqref{wind}} &
\TAc{Ref.~\cite{wang2025topological}} &
\TAc{Descriptor} \\

\addlinespace[2pt]

\TAr{Hedgehog} &
\TAc{Ref.~\cite{bera2025hedgehogtopologicaldefects3d}} &
\TAc{Eigenvector/\\Displacement} &
\TBc{$3$D} &
\TBc{$0$D (point)} &
\TCl{Monopole charge} &
\TAc{---} &
\TAc{Predictor} \\

\addlinespace[5pt]

\rowcolor{gray!12}
\multicolumn{8}{@{}l}{\strut\textbf{Defect densities}} \\

\TAr{Dislocation-like defect} &
\TAc{Ref.~\cite{PhysRevE.109.L053002}} &
\TAc{Displacement\\vector field} &
\TBc{$2$D} &
\TBc{Not applicable} &
\TCl{Continuous Burgers vector, Eq.~\eqref{burg}} &
\TAc{Ref.~\cite{Liu:ib5129}} &
\TAc{Descriptor} \\

\addlinespace[2pt]

\TAr{Geometric charges} &
\TAc{Ref.~\cite{PhysRevE.104.024904}} &
\TAc{Strain\\tensor field} &
\TBc{$2$D/$3$D} &
\TBc{Not applicable} &
\TCl{Continuous geometric charges} &
\TAc{Ref.~\cite{MONDAL2022112609}} &
\TAc{Descriptor} \\

\addlinespace[2pt]

\TAr{Crystal-like densities} &
\TAc{Ref.~\cite{Wang2026UnifiedGeometricPlasticity}} &
\TAc{Displacement\\vector field} &
\TBc{$2$D/$3$D} &
\TBc{Not applicable} &
\TCl{Continuous defect densities} &
\TAc{Ref.~\cite{Wang2026UnifiedGeometricPlasticity}} &
\TAc{Descriptor} \\

\addlinespace[2pt]

\TAr{Nye-like density} &
\TAc{Ref.~\cite{huang2025geometric}} &
\TAc{Eigenvector\\field} &
\TBc{$2$D/$3$D} &
\TBc{Not applicable} &
\TCl{Geometric Nye density} &
\TAc{---} &
\TAc{Predictor} \\

\bottomrule
\end{tabular}
}

\caption{{\color{black}\textbf{Classification of defects in amorphous solids.}}}
\label{tab:defect-classification}
\end{table*}

It is important to emphasize that, at present, the geometric formulation of defects introduced by Moshe and collaborators~\cite{doi:10.1073/pnas.1506531112} arguably constitutes the only fully developed and internally consistent theoretical framework available. Within this approach, defects are defined as geometric charges that quantify deviations of the material’s intrinsic metric from a Euclidean (defect-free) reference metric. They act as sources of internal stress generated by elastic incompatibilities, and their definition is entirely independent of the underlying atomic-scale structure.

Through a multipole expansion, these defects can be systematically classified as monopoles (disclinations), dipoles (dislocations), and quadrupoles (Stone–Wales–type defects). The latter, in particular, appear closely connected to the elastic fields typically associated with shear transformation zones (STZs). At present, however, the precise relationship between the vortex-like topological defects introduced in \cite{wu2023topology} and the geometric charges of Moshe \textit{et al.}~\cite{doi:10.1073/pnas.1506531112,PhysRevE.107.055004} remains unclear. By contrast, the notion of a continuous Burgers vector may be naturally incorporated into this geometric framework, where it would correspond to a dipole field.

It is also worth highlighting that Zhang and collaborators~\cite{wang2020connecting} have proposed an interpretation of micro–shear bands in granular systems in terms of shear-induced rotational symmetry breaking. This opens an intriguing line of thought, as it suggests an underlying symmetry structure that could potentially be formalized into a rigorous order parameter for glasses.

\section*{Discussion}

Given the recent proliferation of approaches using topology to define
defects in amorphous solids, it is useful to summarize these efforts and their main
characteristics in Table~\ref{tab:defect-classification}.
It is important to note that defects in the displacement field are associated with
the actual deformation and rearrangement of the system, whereas defects identified in
eigenvector fields are better viewed as geometric or spectral indices characterizing
the spatial structure of a mode, rather than as objects that themselves undergo a
physical dynamics.

We distinguish between discrete defects, which carry a quantized charge, and continuous
defect densities, which are spatially distributed fields instead of
isolated singularities. We further indicate whether the quantity is a predictor or a
descriptor. Predictors can be identified from a single static configuration, prior to
any deformation, whereas descriptors are defined dynamically and can only be extracted
from the system's deformation or time evolution.

We note that so far many studies on TDs have focused on two-dimensional systems, where
the theoretically relevant quantities are simpler to define and can be more easily
measured in experiments and simulations. The three-dimensional case is considerably
richer and more challenging. Different analyses have identified point-like defects, such as the hedgehog monopoles introduced in
Ref.~\cite{bera2025hedgehogtopologicaldefects3d}, and line defects, such as the vortex
lines considered in Ref.~\cite{wu2024geometrytopologicaldefectsglasses}. At present, it
remains unclear which description is more useful or physically relevant for amorphous
solids. The three-dimensional setting also poses substantial practical difficulties,
since one must resolve particle positions in samples much larger than the relevant
defect length scale, estimated at $20$--$50$ particle
diameters~\cite{wu2024geometrytopologicaldefectsglasses}. Finally, the results of
Ref.~\cite{bera2025hedgehogtopologicaldefects3d} indicate that the charge alone may not
fully characterize a defect in three dimensions: configurations carrying the same charge
can display different spatial structures around their cores, and this geometry appears
to be important for their correlation with local plastic activity.

A further question is whether the defects discussed in this \textit{Perspective}, and
more generally in the recent literature, are genuinely topological in the conventional
sense, or whether the terminology should be used with some caution.
To clarify this point, it is useful to distinguish between different ways in which a
defect can acquire a topological character. Consider first a two-dimensional vector
field $\vec{v}$. Away from points where its amplitude vanishes, it can be written as
$\vec{v}=|\vec{v}|(\cos\theta,\sin\theta),$
so that an integer winding number can be associated with any closed contour on which
$|\vec{v}|\neq0$. The existence of such an integer winding is already a non-trivial
property of the field, but its physical meaning depends on the origin of that winding.

The clearest example is provided by the XY model and by a superfluid. In the XY model,
the magnitude of the order parameter is fixed by construction, whereas in a superfluid
spontaneous symmetry breaking selects a low-energy state with essentially fixed
amplitude and a free phase. In both cases the relevant order-parameter manifold is
$S^1$, with $\pi_1(S^1)=\mathbb{Z}$. 
Here the circle is not merely obtained by normalizing a generic vector field: it is the
physical manifold of the ordered state itself. A non-zero winding therefore cannot be
removed by a smooth deformation that remains within this manifold. Vortices can move,
and vortices of opposite charge can annihilate, but an isolated winding cannot simply
disappear locally. This is the origin of their conventional topological protection.
Together with the stiffness of the ordered phase, this topology also gives rise to the
characteristic energetics, interactions, and collective phenomena associated with
vortices.

For a generic field, whether describing velocity, displacement, or eigenvector, the situation is different. There is no broken-symmetry order parameter fixing the
magnitude of the field, and zeros are naturally allowed. Nevertheless, an isolated zero
can still carry a well-defined integer index, given by the winding of the normalized field around
it. This index is robust under smooth deformations as long as the field remains non-zero
on the surrounding contour: a zero with non-zero index cannot simply disappear in
isolation, although it can move, cross the contour, or annihilate with a zero of opposite
index. The resulting topology is therefore the index topology of singularities of a
vector field, rather than the constrained topology of an order parameter. By itself, such an index
does not imply the long-range energetics, defect interactions, or thermodynamic defect
physics associated with vortices in an ordered phase.

However, it is also true that the vector field describing the displacements in a glass or its eigenvector field are not without important physical constraints. Bonding gives rise to elastic behavior on intermediate and longer length scales. This imposes the constraint of continuity and strong penalties for large gradients. While these do not definitively protect the resulting zeros by constraining them to a mathematically well-defined sub-manifold, they nevertheless give rise to restrictions that may approximate such constraints.

Let us next consider dislocation-like defects and the Burgers vector. Here the important
ingredient is not perfect periodicity of the instantaneous particle positions, but the
existence of an underlying discrete translation structure. In a crystal, continuous
translations are broken down to the discrete lattice $
\Lambda
=
\left\{
n_1\mathbf{a}_1+n_2\mathbf{a}_2
\,\middle|\,
n_1,n_2\in\mathbb{Z}
\right\},$
where $\mathbf{a}_1$ and $\mathbf{a}_2$ are primitive lattice vectors. Displacements
that differ by a lattice vector correspond to the same crystalline order, so that the
order-parameter space is $\mathbb{R}^2/\Lambda\simeq T^2$ with $\pi_1(T^2)=\mathbb{Z}\times\mathbb{Z}$. The Burgers vector is therefore quantized,
$
\mathbf{b}
=
n_1\mathbf{a}_1+n_2\mathbf{a}_2$
where the integers $(n_1,n_2)$ label distinct topological sectors.

Importantly, exact periodicity of the instantaneous particle positions is not required
for this classification. Thermal fluctuations, compositional variations, and smooth elastic deformations may
distort the lattice without destroying its underlying discrete translation structure.
The Burgers vector remains quantized in units of the reference lattice vectors, and the
integers $(n_1,n_2)$ cannot change through a smooth deformation that does not create,
annihilate, or transport a dislocation across the integration contour. It is therefore
the persistence of this discrete translation structure, rather than perfect
instantaneous periodicity, that is essential for the topological classification.

A glass is different since a generic amorphous structure has no discrete translation
subgroup: no finite translation reproduces the same spatial pattern. There is therefore
no analogue of the lattice identification present in a crystal and no corresponding
toroidal order-parameter space. Burgers-like quantities can still
be defined geometrically from the displacement field, but these will take arbitrary
continuous values rather than being quantized. It therefore does not
play the same role as the topological Burgers charge of a crystal. This does not exclude
a broader geometric interpretation of such quantities, but it is important to
distinguish them from the quantized dislocation charge associated with crystalline
translational order.

Taken together, these examples show that the term \textit{topological} can refer to
physically different structures: homotopy classes of an order parameter or indices of
zeros of a continuous field.
Either of these may be mathematically well defined and physically useful, but they need
not imply the same stability, interactions, or collective dynamics.

Beyond these formal distinctions, the more interesting question is how they affect the
physics. Conventional topological defects owe part of their stability to a genuine
topological obstruction, which can lead to conservation laws and characteristic
collective phenomena such as defect pairing, unbinding, and defect-mediated phase
transitions. Defect-like structures in glasses may instead derive their robustness from
a combination of topology, elasticity, geometry, effective interactions, and
dynamical constraints. These may persist even in the absence of crystalline regularity. They may therefore remain long-lived, display characteristic
separations or pairing, and correlate strongly with plastic rearrangements and
mechanical response.
 While the term
\textit{topological} is often used rather broadly, including by us, its precise
meaning should be stated explicitly in each context rather than assumed.

Finally, an important question is what is genuinely gained by formulating amorphous plasticity in terms of topological defects, rather than through established concepts such as soft spots, shear-transformation zones, non-affine response, or local yield criteria. These approaches have provided powerful system-specific indicators of plastic activity, but a broadly transferable structural framework, analogous to the role of dislocations in crystalline plasticity, is still lacking. The topological-defect perspective may offer such a route: it is largely independent of microscopic details, can be applied to both crystalline and amorphous systems within a common language, and is based on physically transparent fields such as displacement or vibrational modes. Whether this framework ultimately provides predictive power beyond existing descriptions remains an open question, but establishing this possibility is one of the main motivations for pursuing the topological viewpoint.

\section*{Open questions}
Before ending, we present a list of questions that we hope will spur progress in applying topological concepts to understanding the properties of amorphous materials.
We have grouped these open questions into four categories: identifying order parameters, formalizing the nature of defects, statistical mechanics of defects, and relating defects to material properties.

\subsection*{Identifying Order Parameters}
\begin{itemize}[noitemsep]
\item Is there a meaningful order parameter in an amorphous solid that allows for a rigorous definition of topological defects?
\begin{itemize}
    \item What is the corresponding non-trivial homotopy group?
    \item Is this order parameter universal?
    \item What is the physical meaning of the phase singularities that have been recently proposed?
\end{itemize}
\item How can the amplitude of the vector field be incorporated into the topological measure?
\item Might multiple order parameters be relevant, and, if so, how might they be related to each other?
\item What are the limits of applying continuum field concepts to inherently discrete systems?
\end{itemize}

\subsection*{Formalizing the Nature of the Defects}
\begin{itemize}[noitemsep]
\item Are topological charges in amorphous solids inherently structural, or do they only relate to response measures such as normal modes and displacements? 
\item While in 2D the ways to define a ``defect'' are relatively limited, in three-dimensional systems several distinct possibilities arise. Is there a notion that is more physically relevant than the others, and if so, why? 
\item How are topological charges defined in different manners related to each other? 
 \item How do these definitions of topological charge evolve as one interpolates between fully amorphous and crystalline limits? 
 \begin{itemize}
 \item Do the newly introduced concepts recover the standard defect structures in crystals \cite{HUANG2025106274}? 
 \item More broadly, is there a unified description of topological defects in crystals and glasses, or are the two fundamentally distinct?
 \item Can the definitions of TD developed for glasses be used to get novel insight into the properties of crystals with defects?
 \end{itemize}
\end{itemize}

\subsection*{Statistical Mechanics of Defects}
\begin{itemize}[noitemsep]
\item Are individual topological defects physically relevant, or must one instead consider coarse-grained topological densities? If so, how should these densities be properly defined and computed?
\item Is it possible to formulate an effective interaction between topological defects? 
\item Can these topological defects be systematically incorporated into mesoscopic elasto-plastic models (see, e.g., \cite{tarjus2026anomalouselasticitymechanicalresponse})? If so, what is the appropriate framework to achieve this?
\item Do thermal fluctuations shift or alter the nature of the defects and their interactions observed in studies of athermal quasistatic shear response? How?
\end{itemize}

\subsection*{Relating Defects to Material Properties}
\begin{itemize}[noitemsep]
   \item How can topological properties be exploited to infer physical behavior? 
   \begin{itemize}
   \item What properties do depend on the presence of topological defects? 
   \item Can TDs tell something about non-linear response?
   \end{itemize}
\item Are all topological defects equally significant? For instance, in two dimensions, do defects with a quadrupolar structure (saddle points) primarily govern the mechanical response? If so, what role is played by the other types of defects?
\item Do topological theories provide new ways of understanding the glass transition \cite{nussinov2018glass,PhysRevE.106.044124,doi:10.1073/pnas.2209144120,PhysRevMaterials.4.125601} and the yielding instability? Do topological structures play a role in strongly supercooled liquids? 
\item Can topological defects be used to formally lay the groundwork for constitutive theories of mechanical response?
\item To what extent are these concepts universal across different interaction potentials, material compositions, and external driving conditions?
\end{itemize}

\section*{Acknowledgments}
We gratefully acknowledge the numerous colleagues and collaborators, past, present, and future for their contributions to this exciting line of research. 
MB acknowledges  the support of the Foreign Young Scholars Research Fund Project (Grant No.22Z033100604) and the sponsorship from the Yangyang Development Fund. MF acknowledges support by the U.S. National Science Foundation (DMREF-2323718).

\end{document}